\documentclass[numeric]{aa}
\usepackage[dvipsnames]{xcolor}

\usepackage{multirow}
\usepackage{graphicx}
\usepackage{txfonts}
\usepackage{hyperref}
\usepackage{xcolor}
\usepackage{subcaption}
\hypersetup{colorlinks,citecolor=blue}
\hypersetup{colorlinks=true,linkcolor=magenta,filecolor=magenta,    urlcolor=blue}

\begin{document} 
\newcommand{\affubb}{Department of Physics, Babeș-Bolyai University, Kogălniceanu Street, Cluj-Napoca, 400084, Romania}
\newcommand{\affunina}{Dipartimento di Fisica ``E. Pancini'', Università degli Studi di Napoli ``Federico II'', Complesso Universitario di Monte Sant’ Angelo, Edificio G, Via Cinthia, I-80126, Napoli, Italy}
\newcommand{\affinfn}{Istituto Nazionale di Fisica Nucleare (INFN), sez. di Napoli, Via Cinthia 9, I-80126 Napoli, Italy}
\newcommand{\affssm}{Scuola Superiore Meridionale, Via Mezzocannone 4, I-80134, Napoli, Italy}
\newcommand{\affiiap}{Indian Institute of Astrophysics, Koramangala II Block, Bangalore 560034, India}
\newcommand{\affunige}{Department of Astronomy, University of Geneva, Chemin Pegasi 51, Versoix, Geneva, 1290, Switzerland}
\newcommand{\affzjnu}{Department of Physics, Zhejiang Normal University, Jinhua 321004, People’s Republic of China}

\title{Evidence of dynamical dark energy found via the DESI DR2 Lyman$\alpha$ forest}

\author{
Salvatore Capozziello\inst{1,2,3} \and
Himanshu Chaudhary\inst{4,5} \and
Ghulam Mustafa\inst{6} \and
S.~K.~J.~Pacif\inst{7}
}

\institute{
$^{1}$\,Dipartimento di Fisica ``E. Pancini'', Universit\`a degli Studi di Napoli ``Federico II'',\\ Complesso Universitario di Monte Sant’ Angelo, Edificio G, Via Cinthia, I-80126 Napoli, Italy \\
$^{2}$\,Istituto Nazionale di Fisica Nucleare (INFN), Sezione di Napoli, Via Cinthia 9, I-80126 Napoli, Italy \\
$^{3}$\,Scuola Superiore Meridionale, via Mezzocannone 4, I-80134 Napoli, Italy \\
$^{4}$\,Department of Physics, Babeș-Bolyai University, Kogălniceanu Street, Cluj-Napoca, 400084, Romania \\
$^{5}$\,Research Center of Astrophysics and Cosmology, Khazar University, 41 Mehseti Street, AZ1096, Baku, Azerbaijan\\ 
$^{6}$\,Department of Physics, Zhejiang Normal University, Jinhua 321004, People’s Republic of China \\
$^{7}$\,Pacif Institute of Cosmology and Selfology (PICS), Sagara, Sambalpur 768224, Odisha, India \\
}

\date{}

\abstract{{We present a comprehensive analysis of the cosmological implications of the Dark Energy Spectroscopic Instrument Data Release 2 Lyman-$\alpha$ forest baryon acoustic oscillation measurements, combined with complementary datasets including DESI DR2 galaxy BAO, Type Ia supernova samples (Pantheon$^+$, DES-Dovekie, and Union3), and the cosmic microwave background CamSpec likelihood. We consider several dark-energy parameterizations such as chevallier–polarski–linder , logarithmic, exponential, jassal–bagla–padmanabhan, barboza–alcaniz, and generalized emergent dark energy, as well as the $w$CDM model and non-flat extensions of the standard $\Lambda$CDM and $w$CDM models. Using the Metropolis-Hastings MCMC algorithm, we constrain the cosmological parameters of each model and compute the Bayesian evidence using the publicly available code \texttt{MCEvidence} to assess the performance of each model relative to $\Lambda$CDM. Our results show that the non-flat extensions remain consistent with spatial flatness, with $\Omega_k \approx 0$ for all dataset combinations. Further, all dark-energy parameterizations predict $w_0 > -1$, $w_a < 0$, and $w_0 + w_a < -1$, which favor a dynamical dark-energy scenario of the Quintom-B type. We also find a moderate preference for dynamical dark-energy models relative to the standard $\Lambda$CDM scenario, reaching up to $\sim3.10\sigma$ for the Ly$\alpha$ + CMB + Galaxy BAO dataset. When DESI DR2 Ly$\alpha$ measurements are combined with different SNe~Ia samples and the CMB, the deviations decrease and remain typically below $\sim2\sigma$, corresponding only to inconclusive preference relative to $\Lambda$CDM. However, this level of evidence is not statistically decisive, and it remains too early to rule out the $\Lambda$CDM model. Finally, the Bayes factor in logarithmic space ($\ln B_{ij}$) shows that model preference relative to $\Lambda$CDM depends strongly on the dataset combination. For Ly$\alpha$ + CMB + Galaxy BAO, $w$CDM and o$w$CDM show moderate evidence, while most other models provide weak or inconclusive evidence. With Pantheon$^{+}$ or DES-Dovekie, o$w$CDM shows strong evidence, whereas other models show moderate evidence.}}

\maketitle

\section{Introduction}\label{sec_1}
The accelerated expansion of the Universe remains one of the most profound mysteries in modern cosmology. In the standard $\Lambda$ cold dark matter ($\Lambda$CDM) paradigm, this acceleration is explained by the cosmological constant $\Lambda$, with a constant equation of state (EoS) parameter $w = -1$. However, recent results from the Dark Energy Spectroscopic Instrument (DESI) Data Release 1 (DR1) \citep{DESI:2024mwx} show deviations from the $\Lambda$CDM paradigm at the levels of 2.6$\sigma$, 2.5$\sigma$, 3.5$\sigma$, and 3.9$\sigma$ when combined with data from the cosmic microwave background (CMB), Pantheon$^+$, Union3, and Dark Energy Survey (DES) 5-year sample of Type Ia supernovae (DES-SN5YR) supernova samples, respectively. Furthermore, the more recent DESI Data Release 2 (DR2) \citep{karim2025desi}, when combined with Pantheon$^+$, Union3, and DES-SN5YR, reaches deviations of 2.8$\sigma$, 3.8$\sigma$, and 4.2$\sigma$, respectively, adding to the available evidence of a dynamical form of dark energy (DE) beyond the simple cosmological constant.

The Lyman-$\alpha$ (Ly$\alpha$) forest serves as one of the most powerful probes of the high-redshift Universe, providing precise constraints on the cosmic expansion rate in the $2 < z < 4$ range \citep{adame2025desi,cuceu2023constraints}. It arises from a series of absorption features in the spectra of distant quasars, produced by neutral hydrogen in the intergalactic medium (IGM). By analysing these absorption patterns along numerous lines of sight, the Ly$\alpha$ forest effectively traces the large-scale structure of matter at early cosmic times. Over the past decade, surveys such as Baryon Oscillation Spectroscopic Survey (BOSS), the extended Baryon Oscillation Spectroscopic Survey (eBOSS), and now DESI have significantly improved measurements of the baryon acoustic oscillation (BAO) scale using both the 3D auto-correlation of Ly$\alpha$ flux and its cross-correlation with quasar positions \citep{delubac2013baryon,slosar2013measurement,des2020completed,adame2025desi}. These studies have demonstrated that the Ly$\alpha$ forest not only provides an independent and robust measurement of the expansion history it also contains additional cosmological information beyond the BAO signal, offering a unique opportunity to test models of DE and possible deviations from the $\Lambda$CDM paradigm.

\citep{cuceu2025desi} present cosmological constraints based on the DESI DR1 Ly$\alpha$ forest, providing the first joint measurements that combine the broadband Alcock-Paczynski (AP) and BAO information at high redshifts, which is referred to as the Best-Ly$\alpha$ sample. When combined, the Best-Ly$\alpha$ sample and the Pantheon$^+$, Union3, and DES-SN5YR supernova samples show a preference for a dynamical DE model at the 0.8$\sigma$, 1.9$\sigma$, and 2.1$\sigma$ levels, respectively. Including CMB data alongside these combinations slightly increases the tension, which reaches 1.1$\sigma$, 2.0$\sigma$, and 2.4$\sigma$ for the same datasets. The deviation becomes more significant when the Ly$\alpha$-AP measurements are combined with the CMB, galaxy BAO, and different Type Ia supernova (SN~Ia) samples, reaching 2.8$\sigma$, 3.8$\sigma$, and 4.2$\sigma$ when adding Pantheon$^+$, Union3, and DES-SN5YR, respectively. In addition, the joint Ly$\alpha$ + galaxy BAO + CMB dataset shows a deviation of about 3.1$\sigma$, while the combination of Ly$\alpha$ + galaxy BAO alone shows a deviation of about 1.6$\sigma$. These findings align with a growing body of recent studies exploring the implications of DESI observations for new physics in the DE sector and potential resolutions to current cosmological tensions. Several recent works have also addressed this issue, aiming to interpret DESI results within a broader theoretical framework and to assess whether they indicate that the standard $\Lambda$ model requires modifications \citep{Odintsov:2025jfq, Odintsov:2024woi,Capozziello:2023ewq,Bernardo:2021cxi,CosmoVerseNetwork:2025alb,alam2025beyond,dinda2025calibration,liu2025torsion,choudhury2024updated,choudhury2025cosmology,choudhury2025cosmological,lee2025constraining,vagnozzi2020new,vagnozzi2023seven,jiang2024nonparametric,pedrotti2025bao,Colgain:2021pmf,Colgain:2025fct,Colgain:2025nzf,Demianski:2018nov,hogaas2025bimetric}.

In this study, we investigated the cosmological implications of the DESI DR2 Ly$\alpha$ forest measurements and explored possible deviations from the standard $\Lambda$ paradigm. To achieve this, we considered several DE models that allow for a time-dependent EoS, and performed a Markov Chain Monte Carlo (MCMC) analysis to constrain the parameter space of each model. We also carried out a statistical analysis to test the consistency of these models with the DESI Ly$\alpha$ data. The manuscript is organized as follows. In Sect.~\ref{sec_2}, we present the mathematical formulation of each DE model. In Sect.~\ref{sec_3}, we describe the dataset used in this analysis and the methodology applied. In Sect.~\ref{sec_4}, we discuss our results and their implications. Finally, in Sect.~\ref{sec_5}, we summarize our main conclusions and outline possible directions for future research.

\section{Standard cosmological background and the $\Lambda$CDM framework}\label{sec_2}
General relativity forms the cornerstone of modern cosmology and satisfies the criteria established by the Lovelock theorem \citep{lovelock1971einstein}, which states that the Einstein field equations are the only second-order field equations derivable from a scalar density in a 4D spacetime. The gravitational dynamics of the concordance model, $\Lambda$CDM, can thus be obtained from the Einstein--Hilbert action, which in natural units ($c = \hbar = 1$) takes the form
\begin{equation}\label{action}
    \mathcal{S} = \frac{1}{16\pi G} \int d^4x \sqrt{-g}\,(R - 2\Lambda) + \mathcal{S}_m,
\end{equation}
where $\Lambda$ is the cosmological constant; $\mathcal{S}_m(g_{\mu\nu},\Psi_m)$ is the action associated with the matter fields, $\Psi_m$; $g$ is the determinant of the spacetime metric, $g_{\mu\nu}$; and $G$ is Newton’s gravitational constant.

By varying the action in Eq.~\eqref{action} with respect to the metric tensor, one obtains the Einstein field equation:
\begin{equation}\label{efe}
    R_{\mu\nu} - \frac{1}{2} R g_{\mu\nu} = 8\pi G \left(T_{\mu\nu} - \frac{\Lambda}{8\pi G} g_{\mu\nu}\right),
\end{equation}
where $R_{\mu\nu}$ and $R$ are the Ricci tensor and scalar, respectively, and $T_{\mu\nu}$ denotes the energy--momentum tensor. For a perfect fluid,
\begin{equation}
    T_{\mu\nu} = (\rho + p)u_\mu u_\nu + p\, g_{\mu\nu},
\end{equation}
with $\rho$ and $p$ being the energy density and pressure, and $u_\mu$ is the four-velocity, satisfying $u_\mu u^\mu = -1$.

The Bianchi identities, $\nabla^\mu G_{\mu\nu} = 0$, together with the Einstein equations, lead to $\nabla^\mu T_{\mu\nu} = 0$, provided $\Lambda$ is constant in spacetime. This ensures the local conservation of energy and momentum for all matter fields.

\subsection{Cosmological dynamics in a flat FLRW background}
Assuming a homogeneous and isotropic universe, the spacetime geometry is described by the flat Friedmann--Lemaître-Robertson-Walker (FLRW) metric,
\begin{equation}\label{metric}
    ds^2 = -dt^2 + a^2(t)\left(dr^2 + r^2 d\Omega^2\right),
\end{equation}
where $a(t)$ denotes the scale factor and $d\Omega^2 = d\theta^2 + \sin^2\theta\, d\phi^2$. Substituting this metric into Eq.~\eqref{efe} yields the Friedmann equations:
\begin{align}
    3H^2 &= 8\pi G \sum_i \rho_i, \label{eq2.1.2}\\
    2\dot{H} + 3H^2 &= -8\pi G \sum_i p_i. \label{eq2.1.3}
\end{align}
Here, $H = \dot{a}/a$ is the Hubble parameter and $\rho_i$ and $p_i$ are the energy density and pressure of the $i$th cosmic component, respectively.

The conservation of the total energy-momentum tensor, $\nabla^\mu T_{\mu\nu}^{\text{(tot)}} = 0$, leads to the continuity equation
\begin{equation}\label{eq2.1.4}
    \dot{\rho}_i + 3H(1 + w_i)\rho_i = 0,
\end{equation}
where $w_i = p_i/\rho_i$ is the EoS parameter.  

For non-relativistic matter ($w_m = 0$), Eq.~\eqref{eq2.1.4} gives $\rho_m = \rho_{m0} a^{-3}$, while for DE characterized by $w_{\text{DE}} = -1$, the density remains constant: $\rho_{\text{DE}} = \rho_{\text{DE},0}$. Consequently, the dimensionless Hubble function,
\begin{equation}
E(z) \equiv \frac{H^2(z)}{H_0^2},
\end{equation}
is expressed as
\begin{equation}\label{eq_8}
\begin{split}
E(z) = & \Omega_m (1+z)^3 + \Omega_{r} (1+z)^4 + \Omega_k (1+z)^2 + \Omega_{\text{DE}} f_{\text{DE}}(z),
\end{split}
\end{equation}
where $\Omega_i = (8\pi G\rho_{i0})/(3H_0^2)$ denotes the present density parameter of each component.

\subsection{Dark energy parametrizations}
While the $\Lambda$CDM model assumes a constant vacuum energy ($w=-1$), theoretical and observational considerations suggest that there are scenarios in which the DE EoS varies with time. For a general EoS $w(z)$, the DE density evolves according to
\begin{equation}\label{eq_9}
f_{\mathrm{DE}}(z) = \exp\!\left[ 3 \int_0^z \frac{1 + w(z')}{1 + z'} \, dz' \right].
\end{equation}
If $w$ is constant, Eq.~\eqref{eq_9} yields $f_{\mathrm{DE}}(z) = (1+z)^{3(1+w)}$, i.e. the $w$ cold dark matter ($w$CDM) model.

To capture possible time evolution, we considered the following EoS parametrizations:
\begin{itemize}
    \item Chevallier-Polarski-Linder (CPL) \citep{chevallier2001accelerating,linder2003exploring}:
    \[
    w(z) = w_0 + w_a \frac{z}{1+z}
    \]
    \item The logarithmic model \citep{efstathiou1999constraining,silva2012thermodynamics}:
    \[
    w(z) = w_0 + w_a \ln(1+z)
    \]
    \item The exponential model \citep{najafi2024dynamical}:
    \[
    w(z) = w_0 + w_a \!\left( e^{\tfrac{z}{1+z}} - 1 \right)
    \]
    \item Jassal-Bagla-Padmanabhan (JBP) \citep{jassal2005wmap}:
    \[
    w(z) = w_0 + w_a \frac{z}{(1+z)^2}
    \]
    \item Barboza-Alcaniz (BA) \citep{barboza2008parametric}:
    \[
    w(z) = w_0 + w_a \frac{z(1+z)}{1+z^2}
    \]
    \item Generalized Emergent DE (GEDE) \citep{li2020evidence}:
    \[
    w(z) = -1 - \frac{\Delta}{3\ln 10} \left[ 1 + \tanh\!\left( \Delta \log_{10}\!\frac{1+z}{1+z_t} \right) \right],
    \]
   where $\Delta$ is the free parameter and $z_t$ is a derived parameter, denoting the redshift at which the matter and DE densities become equal, $\rho_m(z_t)=\rho_{\rm DE}(z_t)$, implying $\Omega_{m0}(1+z_t)^3=\Omega_{\rm DE}(z_t)$.
\end{itemize}
Given any of the $w(z)$ above, $f_{\mathrm{DE}}(z)$ follows from Eq.~\eqref{eq_9}; inserting this into Eq.~\eqref{eq_8} yields the corresponding expansion history, $E(z)$.

\subsection{Non-flat universe extensions}
Although the $\Lambda$CDM model assumes spatial flatness, several recent analyses suggest that the cosmic curvature may deviate slightly from zero. Planck data indicate a mild preference for a closed universe \citep{aghanim2020planck,handley2021curvature,di2020planck}, while low-redshift probes lean towards an open geometry \citep{wu2024measuring}. These tensions motivated us to consider non-flat cosmologies.

\subsubsection{Non-flat $\Lambda$CDM}
Incorporating a curvature component, the dimensionless expansion rate becomes
\begin{equation}\label{eq2.1.6a}
\begin{split}
E^2(z) = \Omega_m (1+z)^3 + \Omega_k (1+z)^2 + \Omega_{\text{DE}},
\end{split}
\end{equation}
where $\Omega_k$ quantifies the present curvature contribution.

\subsubsection{Non-flat $w$CDM}
If DE has a constant EoS of $w \neq -1$, the corresponding expansion rate generalizes to
\begin{equation}\label{eq2.1.6b}
E^2(z) = \Omega_m (1+z)^3 + \Omega_k (1+z)^2 + \Omega_{\text{DE}}(1+z)^{3(1+w)}.
\end{equation}
This model highlights the degeneracy between curvature and the DE EoS parameter, emphasizing the importance of including $\Omega_k$ when constraining cosmic acceleration.

\section{Dataset and methodology}\label{sec_3}
We performed parameter estimation and obtained observational constraints on the free parameters of each DE model proposed in Section~\ref{sec_2}. We now discuss the implications of these models further. To explore the parameter space, we used the Metropolis–Hastings MCMC sampling~\citep{hastings1970monte}, integrated with the cosmological inference code \texttt{Cobaya}~\citep{torrado2021cobaya}. The theoretical models were computed using the Einstein–Boltzmann solver \texttt{CAMB} \citep{lewis2000efficient}. The convergence of the MCMC chains was tested using the Gelman-Rubin statistic $(R - 1)$~\citep{gelman1992inference}, adopting $R - 1 < 0.01$ as the threshold for convergence. The MCMC results were subsequently analysed and visualized using the \texttt{GetDist} package~\citep{lewis2025getdist}.

We applied the Bayesian evidence selection criterion using the publicly available code \texttt{MCEvidence}~\citep{heavens2017marginal,heavens2017no}, which provides an estimate of the marginal likelihood for each cosmological model given the observational data. This allows a direct comparison between two competing models ($i$ and $j$) through the Bayes factor defined as $B_{ij} \equiv Z_i/Z_j$ or, equivalently, in logarithmic form: $\ln B_{ij} = \ln Z_i - \ln Z_j$. A positive value of $\ln B_{ij}$ indicates a preference for model $i$ over model $j$.

To interpret the strength of the evidence, we adopted the revised Jeffreys scale~\citep{kass1995bayes,trotta2008bayes}. According to this criterion, $|\ln B_{ij}| < 1$ corresponds to inconclusive evidence, $1 \leq |\ln B_{ij}| < 2.5$ indicates weak evidence, $2.5 \leq |\ln B_{ij}| < 5$ corresponds to moderate evidence, $5 \leq |\ln B_{ij}| < 10$ indicates strong evidence, and $|\ln B_{ij}| \geq 10$ is considered decisive evidence in favour of one model over the other.

For parameter estimation, we used the Ly$\alpha$ BAO measurements from DESI DR2 and the galaxy BAO measurements from DR2, the SN~Ia sample, and the CamSpec CMB likelihood, which are listed below.
 \begin{itemize}
 \item Ly$\alpha$ forest BAO. We used the Ly$\alpha$ forest BAO measurement from the DR2~\citep{karim2025desi}, which provides constraints on the distance ratios $D_{\mathrm{H}}(z_{\mathrm{eff}})/r_{\mathrm{d}} = 8.63 \pm 0.10$ and  $D_{\mathrm{M}}(z_{\mathrm{eff}})/r_{\mathrm{d}} = 38.98 \pm 0.53$ at the effective redshift $z_{\mathrm{eff}} = 2.33$, with a correlation coefficient of $\rho = -0.43$.

   \item DESI DR2 galaxy BAO. In addition to the Ly$\alpha$ forest BAO measurement, we also used the galaxy clustering BAO data from DESI DR2~\citep{karim2025desi}. This sample provides measurements of $D_{\mathrm{M}}/r_{\mathrm{d}}$ and $D_{\mathrm{H}}/r_{\mathrm{d}}$ in five redshift bins covering the $0.5 < z < 1.5$ range, as well as a single isotropic BAO constraint at $z = 0.295$.

   \item Type Ia supernovae: We also used the unanchored SN~Ia sample from \citep{brout2022pantheon}, which consists of 1,701 light curves from 1,550 SNe~Ia. In our analysis, we excluded supernovae with redshifts of $z < 0.01$ as such low-redshift data are affected by significant systematic uncertainties arising from peculiar velocities. Then, we used the recalibrated 1,820 photometric light curves from DES-Dovekie~\citep{popovic2025dark}, which includes 1,623 DES-discovered SN~Ia and 197 externally sourced low-$z$ supernovae from the Center for Astrophysics (CfA) and Carnegie Supernova Project (CSP) samples~\citep{hicken2009cfa3,hicken2012cfa4,foley2017foundation}. The revised DES-Dovekie has 1,718 SNe~Ia that are also included in DES-SN5YR~\citep{abbott2024dark}. We also used the Union3 compilation~\citep{rubin2025union}, which contains 2,087 SNe~Ia, including 1,363 that overlap with Pantheon$^+$, providing complementary redshift coverage.  For all supernova datasets, we marginalized over the absolute magnitude parameter ($M$) to account for calibration uncertainties (see Equations~A9–A12 in~\citep{goliath2001supernovae}).

   \item CMB \textit{CamSpec} likelihood: Finally, we used the temperature ($TT$), polarization ($EE$), and cross-correlation ($TE$) power spectra measured by \textit{Planck}. For the large angular scales ($\ell < 30$), we used the \textit{Commander} and \textit{SimAll} likelihoods, while for smaller scales ($\ell \geq 30$) we used the \textit{CamSpec} likelihood \citep{efstathiou2020evidence,rosenberg2022cmb}. The \textit{CamSpec} is a new likelihood based on the latest \textit{Planck} PR4 Next-generation Planck Iterative Processing Environment (NPIPE) data release, replacing the earlier Plik likelihood derived from the PR3 release. In addition, we used the CMB lensing measurements obtained from the joint analysis of \textit{Planck} and Atacama Cosmology Telescope (ACT) Data Release 6 (DR6) data~\citep{madhavacheril2024atacama}.
\end{itemize}
The priors chosen for these models are summarized in Table~\ref{tab_1}.

\begin{table}[t]
\caption{Priors on the cosmological parameters used in the analysis.}
\label{tab_1}
\centering
\begin{tabular}{llll}
\hline
Parametrization & parameter & default & prior \\
\hline
$\Lambda$CDM 
& $\Omega_{c}h^2$ & --- & $\mathcal{U}[0.001, 0.99]$ \\
& $\Omega_{b}h^2$ & --- & $\mathcal{U}[0.005, 0.1]$ \\
& $100\theta_\mathrm{MC}$ & --- & $\mathcal{U}[0.5, 10]$ \\
& $\ln(10^{10} A_\mathrm{s})$ & --- & $\mathcal{U}[1.61, 3.91]$ \\
& $n_\mathrm{s}$ & --- & $\mathcal{U}[0.8, 1.2]$ \\
& $\tau$ & --- & $\mathcal{U}[0.01, 0.8]$ \\
\hline
Extended 
& $\Omega_k$ & 0 & $\mathcal{U}[-0.3, 0.3]$ \\
\hline
Dark energy 
& $w$ or $w_0$ & $-1$ & $\mathcal{U}[-3, 1]$ \\
& $w_{a}$ & $0$ & $\mathcal{U}[-3, 2]$ \\
\hline
Emergent 
& $\Delta$ & --- & $\mathcal{U}[-3, 10]$ \\
\hline
\end{tabular}
\begin{flushleft}
\textit{Notes.} The symbol $\mathcal{U}$ denotes uniform priors. The parameters listed correspond to the baseline $\Lambda$CDM model and its extensions, including curvature and dynamical DE parametrizations.
\end{flushleft}
\end{table}

\begin{figure*}
\centering
\includegraphics[scale=0.40]{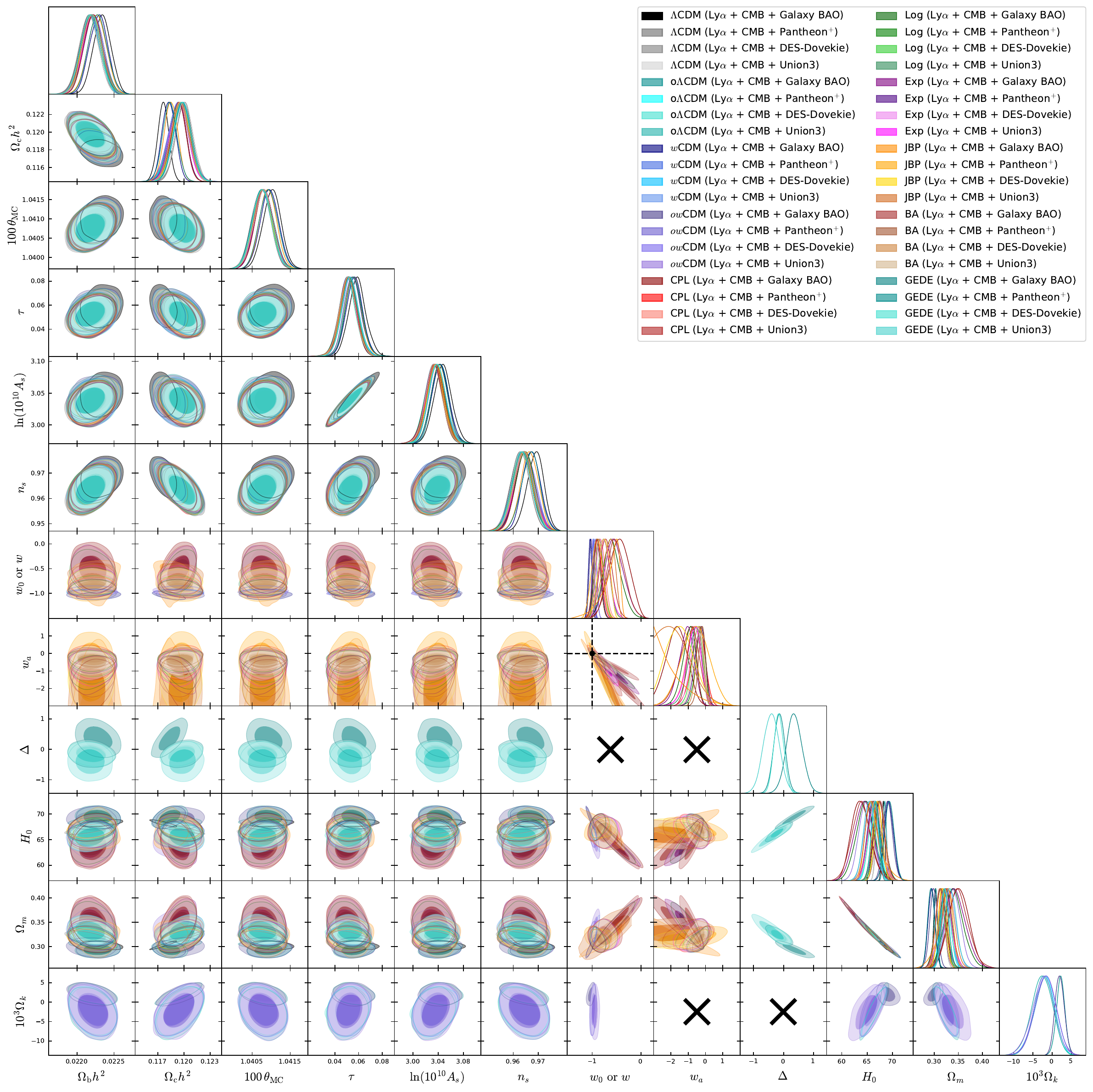}
\caption{Corner plot obtained by superimposing the $\Lambda$CDM, o$\Lambda$CDM, $w$CDM, o$w$CDM, CPL, logarithmic, exponential, JBP, BA, and GEDE models using DESI DR2 Ly$\alpha$ forest measurements in combination with the CMB, galaxy BAO, and SN~Ia (Pantheon$^+$, DES-Dovekie, and Union3) datasets. The contours correspond to the 68\% (1$\sigma$) and 95\% (2$\sigma$) confidence levels. The cross symbols indicate parameters that are not present in the corresponding models.}
\label{fig_0}
\end{figure*}

\begin{figure*}
\begin{subfigure}{.33\textwidth}
\includegraphics[width=\linewidth]{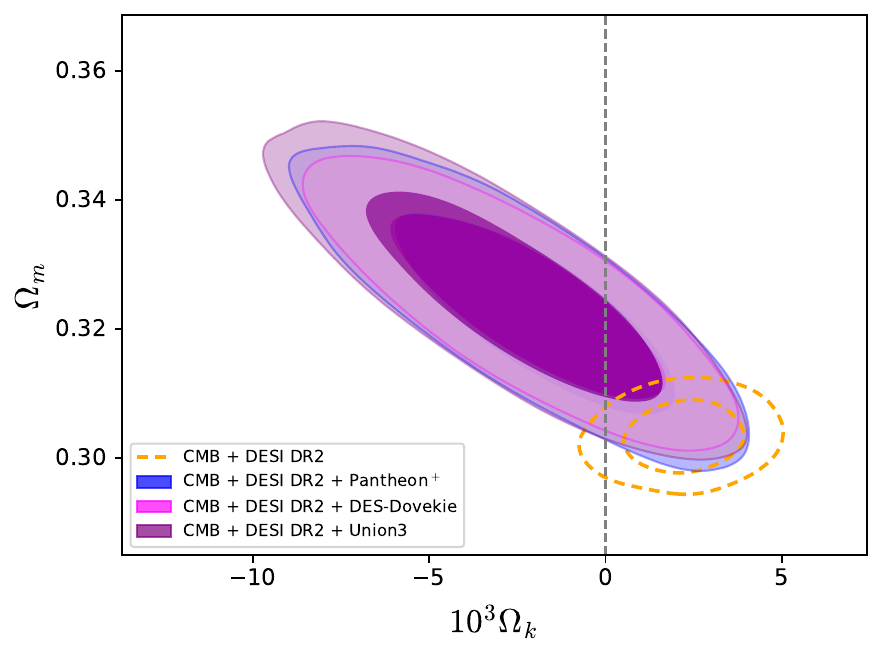}
    \caption{o$\Lambda$CDM}\label{fig_1a}
\end{subfigure}
\hfil
\begin{subfigure}{.33\textwidth}
\includegraphics[width=\linewidth]{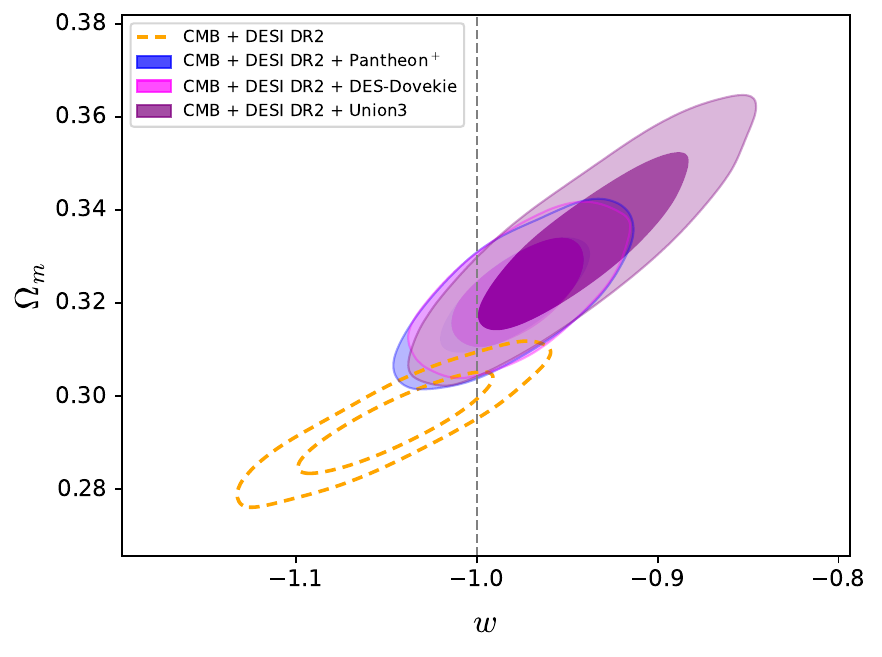}
    \caption{$w$CDM}\label{fig_1b}
\end{subfigure}
\hfil
\begin{subfigure}{.33\textwidth}
\includegraphics[width=\linewidth]{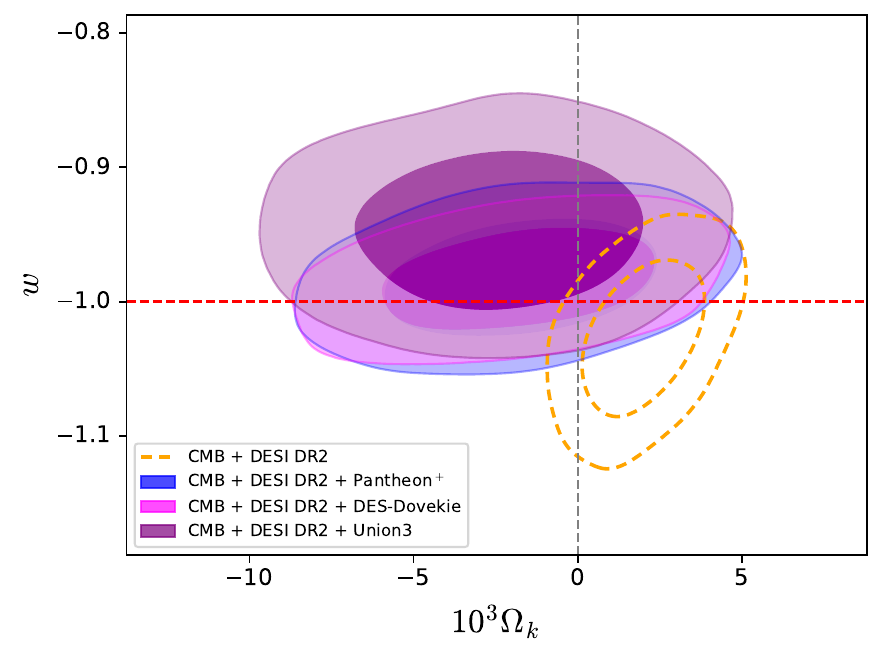}
    \caption{o$w$CDM}\label{fig_1c}
\end{subfigure}
\begin{subfigure}{.33\textwidth}
\includegraphics[width=\linewidth]{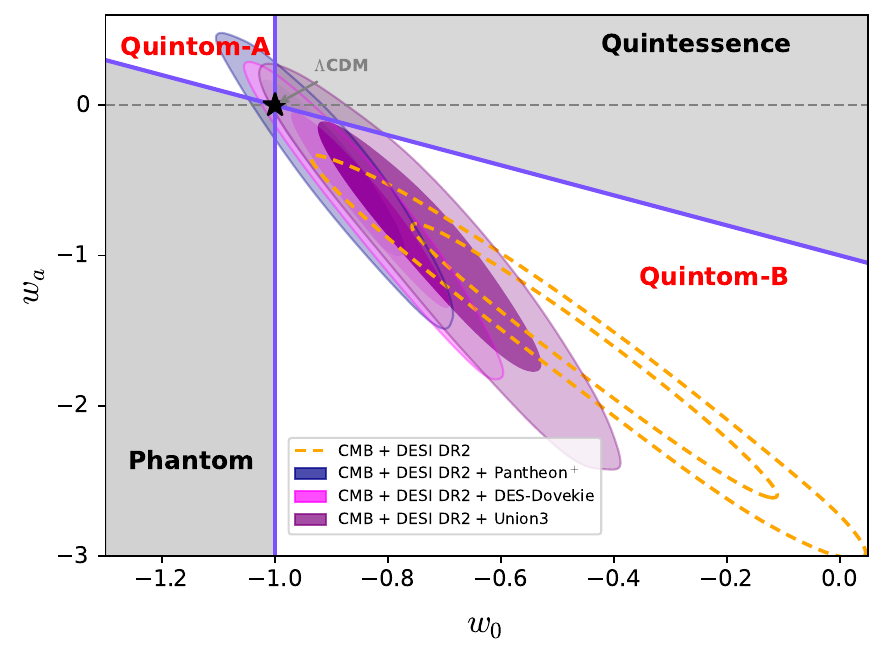}
    \caption{CPL}\label{fig_1d}
\end{subfigure}
\hfil
\begin{subfigure}{.33\textwidth}
\includegraphics[width=\linewidth]{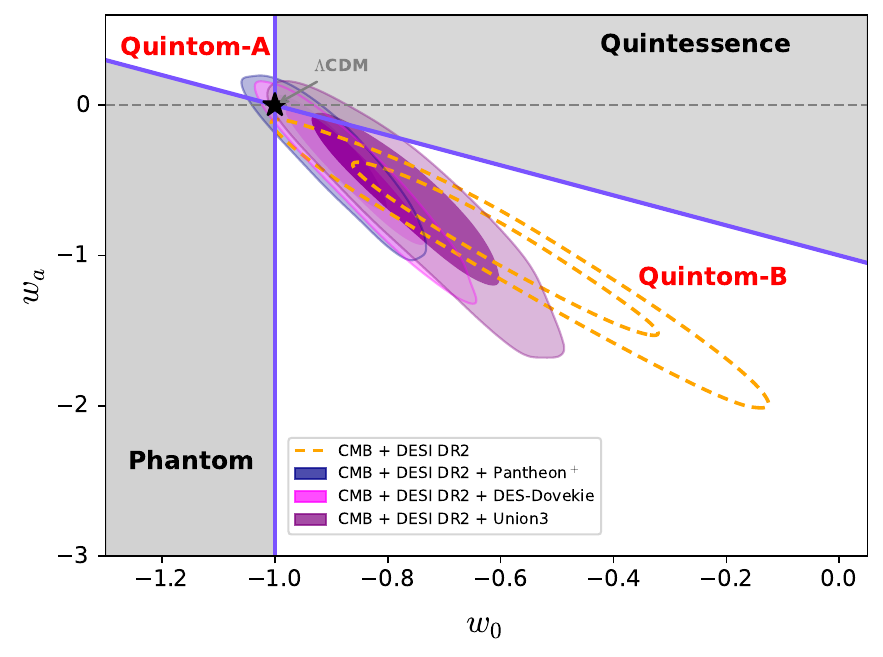}
     \caption{logarithmic}\label{fig_1e}
\end{subfigure}
\hfil
\begin{subfigure}{.33\textwidth}
\includegraphics[width=\linewidth]{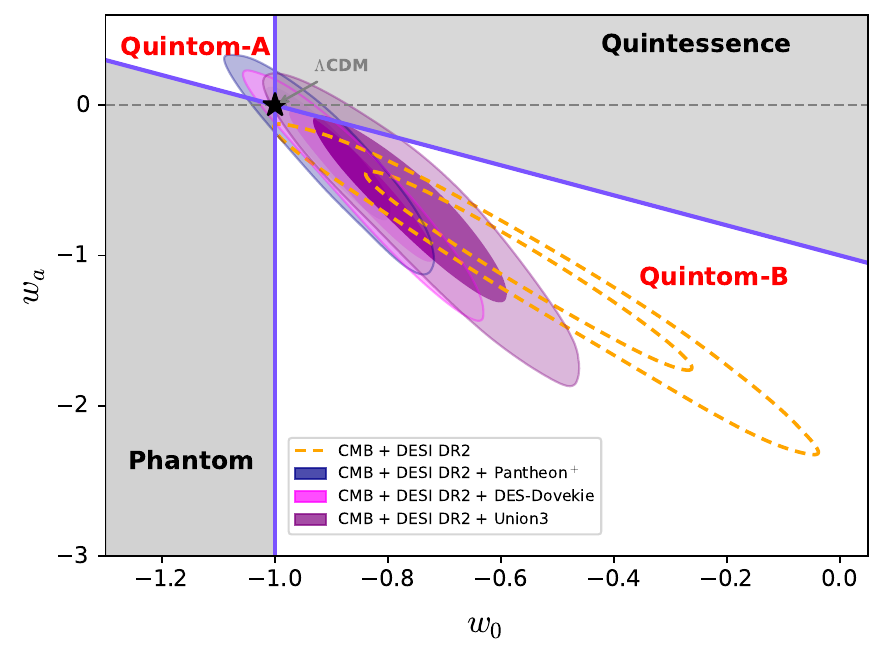}
    \caption{exponential}\label{fig_1f}
\end{subfigure}
\hfil
\begin{subfigure}{.33\textwidth}
\includegraphics[width=\linewidth]{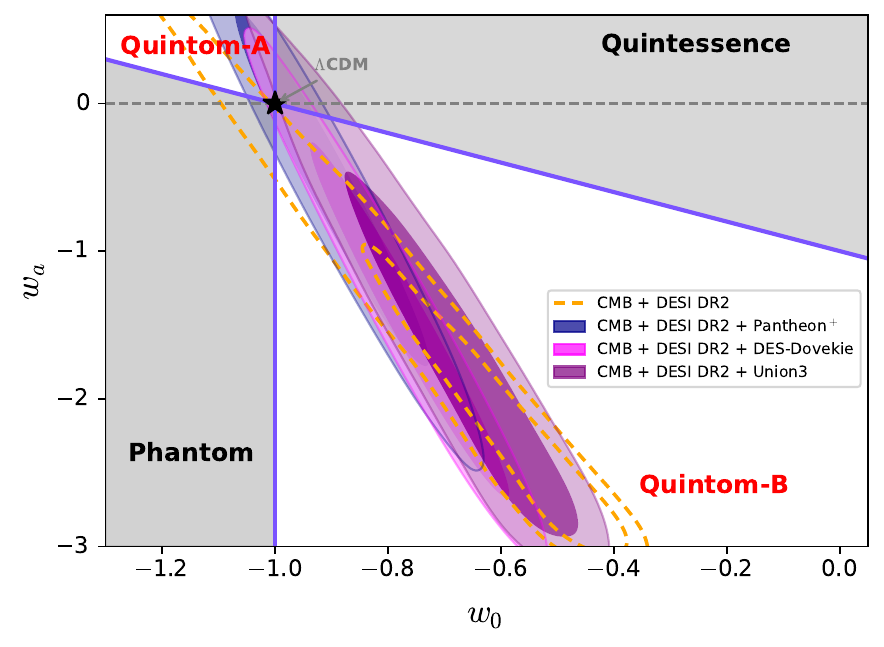}
    \caption{JBP}\label{fig_1g}
\end{subfigure}
\hfil
\begin{subfigure}{.33\textwidth}
\includegraphics[width=\linewidth]{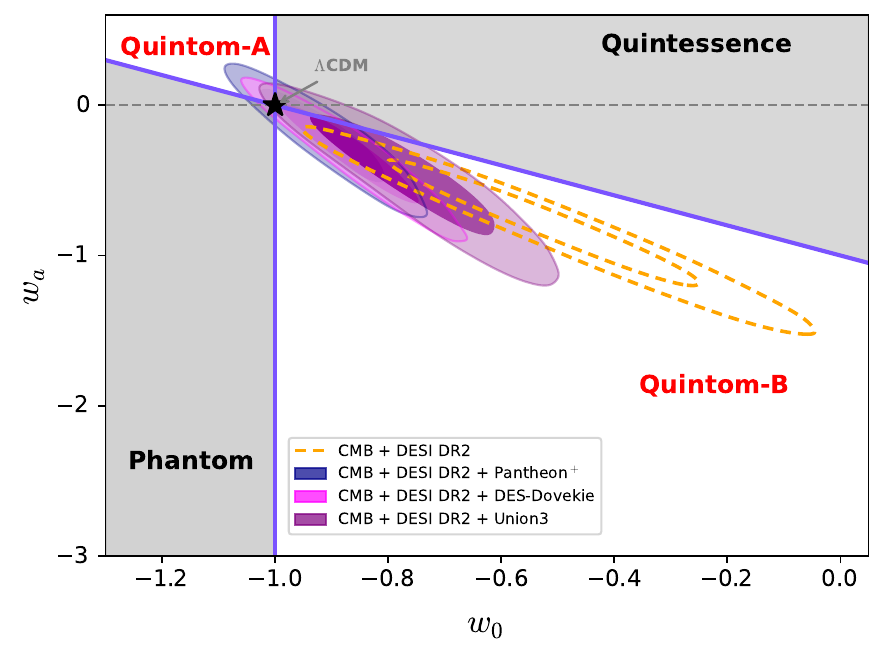}
     \caption{BA}\label{fig_1h}
\end{subfigure}
\hfil
\begin{subfigure}{.33\textwidth}
\includegraphics[width=\linewidth]{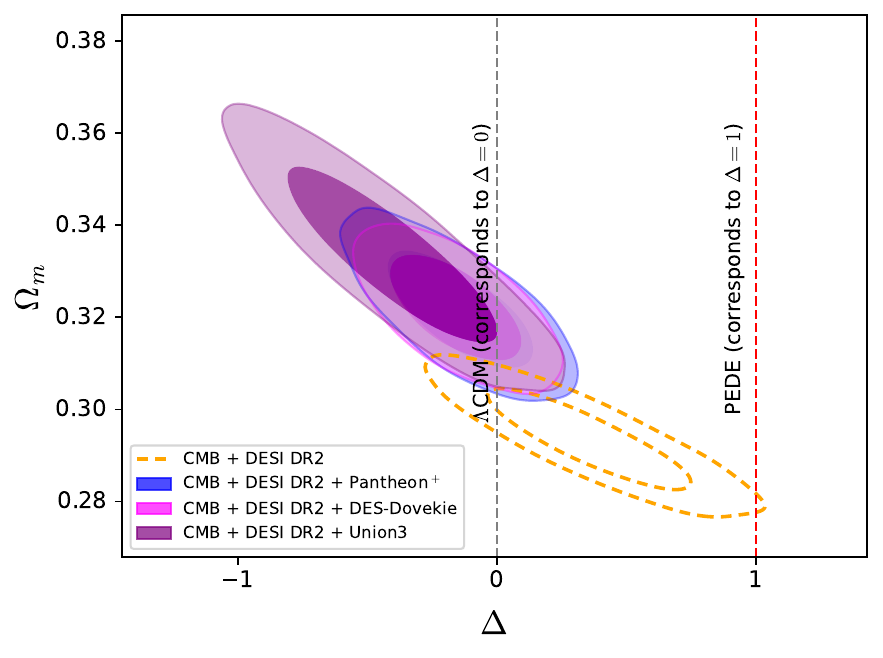}
     \caption{GEDE}\label{fig_1i}
\end{subfigure}
\caption{2D marginalized confidence contours of different planes of the o$\Lambda$CDM, $w$CDM, o$w$CDM, CPL, logarithmic, exponential, JBP, BA, and GEDE models using DESI DR2 Ly$\alpha$ forest measurements in combination with the CMB, galaxy BAO, and SN~Ia (Pantheon$^+$, DES-Dovekie, and Union3) measurements at 68\% (1$\sigma$) and 95\% (2$\sigma$) confidence intervals.}\label{fig_1}
\end{figure*}

\begin{figure*}
\centering
\includegraphics[scale=0.50]{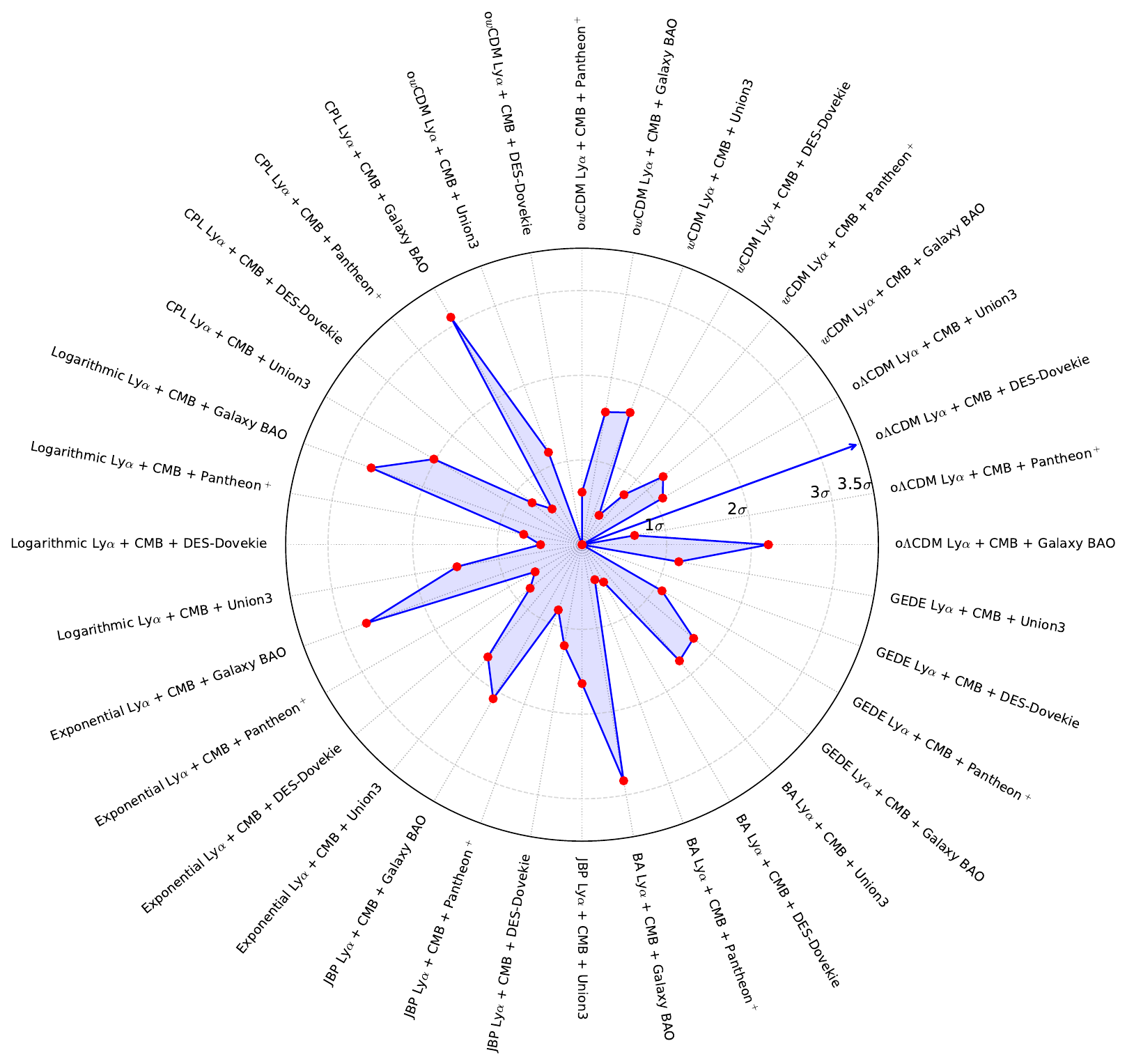}
\caption{Radar plot showing the deviation of the o$\Lambda$CDM, $w$CDM, o$w$CDM, CPL, logarithmic, exponential, JBP, BA, and GEDE models in terms of $\sigma$. Each radial distance, ending at a red dot, represents the statistical significance of the deviation. The radial axes correspond to the different dataset combinations, which are labelled along the outer edge of the plot.}
\label{fig_2}
\end{figure*}

\section{Results}\label{sec_4}
In Fig.~\ref{fig_0}, we show the corner plot obtained by superimposing each model considered in this paper. The off-diagonal panels show the 2D marginalized confidence contours at the 68\% and 95\% confidence levels, while the diagonal panels show the 1D marginalized posterior distributions for each parameter. Fig.~\ref{fig_1} shows the 2D marginalized confidence contours at the $1\sigma$ and $2\sigma$ confidence levels for the different cosmological models. In this analysis, we used the parameter values predicted by the $\Lambda$CDM model for each combination of datasets, treating it as the baseline model for comparison. In cosmology, we rely on observation-based inference, with which it is not possible to repeat measurements under identical conditions to achieve the same degree of precision. For this reason, evidence at the level of 2–4 $\sigma$ is generally considered significant. Several well-known anomalies such as the Hubble tension, the $S_8$ tension, the $M_B$ calibration tension, and the CMB lensing anomaly are discussed as 'tensions' precisely because they appear within this range. Therefore, we quantified the deviation of each model parameter from the $\Lambda$CDM baseline in terms of the tension ($T$), defined as $T = \frac{|x_{\text{model}} - x_{\Lambda \text{CDM}}|}{\sqrt{\sigma_{\text{model}}^2 + \sigma_{\Lambda \text{CDM}}^2}},$ where $x_{\text{model}}$ and $x_{\Lambda \text{CDM}}$ are the predicted parameter values (e.g. $H_0$, $\Omega_m$, etc.), and $\sigma_{\text{model}}$ and $\sigma_{\Lambda \text{CDM}}$ are their uncertainties \citep{camera2019does,chang2019unified}. The results are interpreted using the following scale: $|T| < 1\sigma$, which is consistent with $\Lambda$CDM; $1\sigma \le |T| < 2.5\sigma$, which is consistent with inconclusive tension; $2.5\sigma \le |T| < 5\sigma$, which is consistent with moderate tension; $5\sigma \le |T| < 10\sigma$, which is consistent with strong tension; and $|T| \ge 10\sigma$, which is consistent with decisive tension.

First, we compared the predicted values of $H_0$ for each model with those obtained from the $\Lambda$CDM model across different dataset combinations. The level of tension depends both on the DE parametrization and the dataset considered. For the Ly$\alpha$ + CMB + Galaxy BAO combination, the CPL, logarithmic, exponential, BA, $w$CDM, JBP, and GEDE models show inconclusive evidence of tension with respect to the $\Lambda$CDM prediction for $H_0$, with deviations in the range $1\sigma \lesssim |T| \lesssim 2.5\sigma$. In contrast, the open $\Lambda$ cold dark matter (o$\Lambda$CDM) and open $w$ cold dark matter (o$w$CDM) models remain consistent with $\Lambda$CDM. For the Ly$\alpha$ + CMB + Pantheon$^{+}$ and Ly$\alpha$ + CMB + DES-Dovekie combinations, all considered DE models remain consistent with the $\Lambda$CDM prediction for $H_0$, as the corresponding tension values satisfy $|T| < 1\sigma$. Finally, for Ly$\alpha$ + CMB + Union3, the $w$CDM, o$w$CDM, JBP, and GEDE models show inconclusive evidence of tension with respect to $\Lambda$CDM, whereas the remaining parametrizations remain consistent with the baseline model.

Next, we examined the matter density parameter ,$\Omega_m$, and its deviation from the $\Lambda$CDM model across the different dataset combinations. For the Ly$\alpha$ + CMB + Galaxy BAO combination, the CPL, logarithmic, exponential, BA, $w$CDM, JBP, and GEDE models show inconclusive evidence of tension with respect to the $\Lambda$CDM prediction for $\Omega_m$, with deviations in the range $1\sigma \lesssim |T| \lesssim 2.5\sigma$. In contrast, the o$\Lambda$CDM and o$w$CDM models remain consistent with $\Lambda$CDM. For the Ly$\alpha$ + CMB + Pantheon$^{+}$ and Ly$\alpha$ + CMB + DES-Dovekie combinations, all considered DE models remain statistically consistent with the $\Lambda$CDM prediction for $\Omega_m$, as the corresponding tension values satisfy $|T| < 1\sigma$. Finally, for the Ly$\alpha$ + CMB + Union3 combination, the $w$CDM, o$w$CDM, JBP, and GEDE models exhibit inconclusive evidence of tension with respect to $\Lambda$CDM, whereas the remaining parametrizations remain consistent with the baseline model.

Now, Figs.~\ref{fig_1a}, \ref{fig_1b}, and \ref{fig_1c} show the 2D marginalized confidence contours in the $\{10^3\Omega_k-\Omega_m\}$, $\{w-\Omega_m\}$, and $\{10^3\Omega_k-w\}$ planes for the o$\Lambda$CDM, $w$CDM, and o$w$CDM models, respectively. First, we discuss the constraints on the curvature parameter, $\Omega_k$. Since the numerical values of the curvature parameter are very close to zero, we constrained $\Omega_k$ in the analysis, but we present the results in terms of $10^{3}\Omega_k$ for better visualization. For the o$\Lambda$CDM model, the curvature parameter is predicted to be $10^{3}\Omega_k = 2.2 \pm 1.1$ and $10^{3}\Omega_k = -2.1^{+2.8}_{-2.4}$ for the Ly$\alpha$ + CMB + Galaxy BAO and Ly$\alpha$ + CMB + Pantheon$^{+}$ dataset combinations, respectively. When combined with DES-Dovekie and Union3, the o$\Lambda$CDM model predicts $10^{3}\Omega_k = -2.2^{+2.6}_{-2.4}$ and $10^{3}\Omega_k = -2.7^{+2.9}_{-2.6}$, respectively. Similarly, for the o$w$CDM model, the curvature parameter is predicted to remain close to zero for all dataset combinations, with $10^{3}\Omega_k = 2.0 \pm 1.2$, $10^{3}\Omega_k = -1.8 \pm 2.8$, $10^{3}\Omega_k = -1.9 \pm 2.8$, and $10^{3}\Omega_k = -2.4^{+3.0}_{-2.7}$ for the Ly$\alpha$ + CMB + Galaxy BAO, Ly$\alpha$ + CMB + Pantheon$^{+}$, Ly$\alpha$ + CMB + DES-Dovekie, and Ly$\alpha$ + CMB + Union3 dataset combinations, respectively. All these results remain fully consistent with a spatially flat Universe ($\Omega_k = 0$) and are also in close agreement with previous measurements from Wilkinson Microwave Anisotropy Probe (WMAP) ($-0.0179 < \Omega_k < 0.0081$, 95\% CL) \citep{bennett2013nine}, Balloon Observations Of Millimetric Extragalactic Radiation and Geophysics (BOOMERanG) ($0.988 < \Omega_{M/R} + \Omega_\Lambda < 1.0081$, 95\% CL) \citep{de2000flat}, and Planck ($\Omega_{M/R} + \Omega_\Lambda = 1.00 \pm 0.026$, 68\% CL) \citep{aghanim2020planck}. Further, the $w$CDM model predicts $w = -1.046 \pm 0.036$ when the Ly$\alpha$ data are combined with the CMB and Galaxy BAO samples, whereas when Ly$\alpha$ is combined with the CMB and different SN~Ia samples, it predicts $w > -1$.

Figs.~\ref{fig_1d}, \ref{fig_1e}, \ref{fig_1f}, \ref{fig_1g}, and \ref{fig_1h} show the 2D marginalized confidence contours at the $1\sigma$ and $2\sigma$ confidence levels in the $\{w_0-w_a\}$ plane for the CPL, logarithmic, exponential, JBP, and BA models, respectively. All models predict $w_0 > -1$ and $w_a < 0$ when the Ly$\alpha$ data are combined with the CMB, different SN~Ia samples (Pantheon$^+$, DES-Dovekie, and Union3), and Galaxy BAO, indicating a preference for the DE scenario characterized by $w_0 > -1$, $w_a < 0$, and $w_0 + w_a < -1$, namely the Quintom-B scenario \citep{cai2025quintom,ye2025hints}, in which the EoS parameter satisfies $w < -1$ in the past and evolves to $w > -1$ at the present epoch. By crossing $w = -1$, this behaviour corresponds to a phantom crossing \citep{silva2025testing}.

Fig.~\ref{fig_1i} shows the 2D marginalized confidence contours in the $\{\Delta-\Omega_m\}$ plane for the GEDE model. We find that the GEDE model predicts $\Delta = 0.36\pm0.26$ when the Ly$\alpha$ data are combined with the CMB and Galaxy BAO samples, showing an inconclusive deviation from the $\Lambda$CDM model. However, when different SN~Ia samples (Pantheon$^+$, DES-Dovekie, and Union3) are included, the GEDE model yields negative values of $\Delta$, indicating a possible injection of DE at high redshifts\citep{lodha2025desi,lodha2025extended}.

In Fig.~\ref{fig_2}, we present a radar plot showing the deviations of the o$\Lambda$CDM, $w$CDM, o$w$CDM, CPL, logarithmic, exponential, JBP, BA, and GEDE models relative to the $\Lambda$CDM model. The deviations are quantified using the $N\sigma$ analysis based on the differences in the minimum $\chi^2$ values, which are defined as $\Delta \chi^2_{\rm MAP} \equiv \chi^2_{\rm Model} - \chi^2_{\Lambda{\rm CDM}}$. As our main goal was to assess the preference for dynamical DE over the $\Lambda$CDM model, we focus on the CPL, logarithmic, exponential, JBP, BA, and GEDE models in our further discussion. Since $\Lambda$CDM is a special case of the dynamical DE model, the statistic  $\Delta \chi^2_{\rm MAP}$ is expected to follow a $\chi^2$ distribution, with degrees of freedom equal to the number of additional parameters  in the extended model, assuming Gaussian errors and that the null hypothesis ($\Lambda$CDM) is valid. For easier interpretation, we converted $\Delta\chi^2_{\rm MAP}$ into an equivalent frequentist significance ($N\sigma$) for a 1D Gaussian distribution, $\mathrm{CDF}_{\chi^2}\!\left(\Delta \chi^2_{\rm MAP}\,|\,k~{\rm dof}\right)
=
\frac{1}{\sqrt{2\pi}}
\int_{-N}^{N} e^{-x^2/2}\,dx,$ where $k$ denotes the number of additional parameters in the extended model, and the left-hand side represents the cumulative distribution function of the $\chi^2$ distribution.

For the Ly$\alpha$ + CMB + Galaxy BAO combination, the CPL model shows moderate preference ($\simeq 3.10\sigma$) relative to $\Lambda$CDM, while the logarithmic, exponential, and BA models also show moderate preference at the $\sim 2.6$–$2.8\sigma$ level. The JBP and GEDE models show an inconclusive preference ($\simeq 2.10\sigma$ and $\simeq 1.72\sigma$, respectively) relative to $\Lambda$CDM. For the Ly$\alpha$ + CMB + Pantheon$^+$ dataset, all dynamical DE models remain statistically consistent with $\Lambda$CDM, with deviations close to or slightly above $1\sigma$. The largest deviation arises for the GEDE model ($\simeq 1.09\sigma$), which only corresponds to an inconclusive preference. For the Ly$\alpha$ + CMB + DES-Dovekie combination, all models remain statistically consistent with $\Lambda$CDM, with deviations below $\sim 1.3\sigma$. The JBP model shows the largest deviation ($\simeq 1.21\sigma$), which still lies within the range of inconclusive preference, while the GEDE model shows no deviation from $\Lambda$CDM. Finally, for the Ly$\alpha$ + CMB + Union3 combination, the CPL model shows inconclusive preference ($\simeq 2.02\sigma$); the remaining models also fall within the range of inconclusive preference, with deviations in the $\sim 1.2$–$1.8\sigma$ range relative to $\Lambda$CDM.

Based on the logarithmic form of the Bayes factor ,$|\ln B_{i,j}|$, and following the revised Jeffreys scale, we obtained the following results. For the Ly$\alpha$ + CMB + Galaxy BAO combination, the $w$CDM and o$w$CDM models show moderate evidence against $\Lambda$CDM, while the o$\Lambda$CDM, CPL, JBP, and GEDE models provide weak evidence. In contrast, the logarithmic, exponential, and BA models yield inconclusive evidence relative to the $\Lambda$CDM baseline. For the Ly$\alpha$ + CMB + Pantheon$^{+}$ combination, the o$w$CDM model shows strong evidence against $\Lambda$CDM, whereas the o$\Lambda$CDM, $w$CDM, CPL, logarithmic, exponential, JBP, BA, and GEDE models show moderate evidence. For the Ly$\alpha$ + CMB + DES-Dovekie dataset, the o$w$CDM model again shows strong evidence against $\Lambda$CDM. The o$\Lambda$CDM, $w$CDM, CPL, logarithmic, exponential, BA, and GEDE models provide moderate evidence, while the JBP model shows weak evidence. Finally, for the Ly$\alpha$ + CMB + Union3 combination, the o$\Lambda$CDM and o$w$CDM models provide moderate evidence against $\Lambda$CDM. The $w$CDM, exponential, logarithmic, CPL, BA, JBP, and GEDE models also show moderate evidence relative to the $\Lambda$CDM prediction.

\section{Discussion and conclusions}\label{sec_5}
In this work, we investigated the cosmological implications of the DESI DR2 Ly$\alpha$ forest measurements in combination with several datasets, including the CMB, Galaxy BAO, and multiple SN~Ia compilations (Pantheon$^+$, DES-Dovekie, and Union3), to test the preference of the dynamical DE model over the $\Lambda$CDM model. Using various redshift-dependent DE parametrizations (CPL, logarithmic, exponential, JBP, BA, and GEDE models), as well as constant EoS and non-flat extensions, we performed a comprehensive Bayesian analysis based on the Metropolis–Hastings MCMC algorithm implemented in \texttt{Cobaya}. The theoretical predictions were computed using the Einstein–Boltzmann solver \texttt{CAMB}. Model comparison was carried out using the Bayesian evidence computed with \texttt{MCEvidence}, interpreted through the revised Jeffreys scale. Our main results can be summarized as follows:
\begin{itemize}
\item Hubble and matter density parameters. The inclusion of DESI DR2 Ly$\alpha$ measurements together with other cosmological datasets shows dataset-dependent variations in both the Hubble constant ,$H_0$, and the matter density parameter ,$\Omega_m$. In particular, the Ly$\alpha$ + CMB + Galaxy BAO combination shows the largest deviations from the $\Lambda$CDM predictions. For this dataset combination, several DE parametrizations, including the CPL, logarithmic, exponential, BA, $w$CDM, JBP, and GEDE models, show inconclusive evidence of tension with respect to the $\Lambda$CDM prediction for both $H_0$ and $\Omega_m$, with deviations typically in the $1\sigma \lesssim |T| \lesssim 2.5\sigma$ range. In contrast, the o$\Lambda$CDM and o$w$CDM models remain statistically consistent with $\Lambda$CDM.\\
\item Spatial curvature. The non-flat extensions o$\Lambda$CDM and o$w$CDM yield curvature constraints consistent with $\Omega_k \approx 0$, with all dataset combinations providing tightly constrained values of $\Omega_k$ that are statistically consistent with zero. These findings support the assumption of a spatially flat Universe, in agreement with previous results from WMAP, BOOMERanG, and Planck.\\
\item Quintom-B scenario. All redshift-dependent models (CPL, logarithmic, exponential, JBP, and BA) predict values of $w_0 > -1$ and $w_a < 0$, as indicated by the 2D marginalized confidence contours at the $1\sigma$ and $2\sigma$ confidence levels in the $\{w_0 - w_a\}$ plane. These results show that the DE scenario is characterized by a Quintom-B type ($w_0 > -1$, $w_a < 0$, and $w_0 + w_a < -1$) in which the EoS parameter satisfies $w < -1$ in the past and evolves to $w > -1$ at the present epoch by crossing the phantom divide, $w = -1$, rather than a cosmological constant model in which DE is characterized by $w = -1$.\\
\item Evidence for dynamical DE. The inclusion of Ly$\alpha$ data in the combined cosmological analysis shows that the agreement with the $\Lambda$CDM model depends strongly on the choice of DE parametrization and dataset combination. For the Ly$\alpha$ + CMB + Galaxy BAO dataset, the CPL model shows moderate preference relative to $\Lambda$CDM ($\sim 3.10\sigma$), while the logarithmic, exponential, and BA models also exhibit a moderate preference ($\sim 2.6$–$2.8\sigma$). The JBP and GEDE models only show an inconclusive preference. When Ly$\alpha$ is combined with CMB and Pantheon$^{+}$ or DES-Dovekie, all models remain statistically consistent with $\Lambda$CDM. For the Ly$\alpha$ + CMB + Union3 combination, the CPL model shows only an inconclusive preference, while the remaining parametrizations remain broadly compatible with the $\Lambda$CDM prediction.
\item Bayesian model comparison. Based on the Bayesian evidence and the revised Jeffreys scale, the preference for dynamical DE models relative to $\Lambda$CDM depends strongly on the dataset combination. For the Ly$\alpha$ + CMB + Galaxy BAO dataset, $w$CDM and o$w$CDM show moderate evidence, while most other models provide weak or inconclusive evidence. When combined with Pantheon$^{+}$ or DES-Dovekie, the o$w$CDM model shows strong evidence against $\Lambda$CDM, whereas the remaining models generally yield moderate evidence.
\end{itemize}
Our results indicate that when high-redshift DESI DR2 Ly$\alpha$ forest measurements are combined with other cosmological datasets, the level of agreement with the $\Lambda$CDM model becomes strongly dependent on the adopted DE parametrization and dataset combination. For the Ly$\alpha$ + CMB + Galaxy BAO dataset, deviations from $\Lambda$CDM reach the $\sim1.7$–$3.10\sigma$ level for several dynamical DE models, while the logarithmic, exponential, and BA parametrizations show a moderate preference at the $\sim2.6$–$2.8\sigma$ level. When different SN~Ia samples are included, the deviations decrease significantly and remain typically below $\sim2\sigma$, indicating only an inconclusive preference relative to $\Lambda$CDM. Although these deviations are not statistically decisive, they highlight the sensitivity of cosmological constraints to high-redshift Ly$\alpha$ measurements and suggest that future high-precision observations such as those in DESI Data Release 3 may provide further insight into the nature of DE.

The upcoming Stage IV surveys will shed new light on the nature of DE. DESI will deliver additional DR2 constraints from full-shape fitting, bispectrum, and gravitational lensing in 2025–2026, followed by DR3 BAO results expected in 2027; this will provide new insights into the state of DE, the $\Lambda$CDM model, and the phantom crossing. Observations from the \textit{Hubble} Space Telescope and James Webb Space Telescope will refine $H_0$, while the Simons Observatory~\citep{ade2019simons}, Legacy Survey of Space and Time (LSST) at the \textit{Vera C. Rubin} Observatory~\citep{ade2019simons}, and ESA’s \textit{Euclid} mission~\citep{laureijs2011euclid} will soon provide new CMB, weak lensing, and supernova data. The Subaru Prime Focus Spectrograph (PFS) survey~\citep{takada2014extragalactic} and the \textit{Nancy Grace Roman} Space Telescope~\citep{spergel2015wide} will extend DE constraints beyond $z > 1$, and in the 2030s DESI-II~\citep{dawson2022snowmass2021} will push these boundaries even further. If Stage IV surveys challenge the standard $\Lambda$CDM model, the next questions will concern how we move forward, which new observing methods or cross-survey approaches we should focus on first, and whether there any particular signals or patterns in the data that could point us towards the true nature of DE. These will not be easy questions to answer. In the long run, we may find that proving $\Lambda$CDM is incomplete was the simple part and that understanding what actually drives cosmic acceleration is a much harder task.

\begin{acknowledgements}
SC acknowledges the Istituto Nazionale di Fisica Nucleare (INFN) Sez. di Napoli,  Iniziative Specifiche QGSKY and MoonLight-2  and the Istituto Nazionale di Alta Matematica (INdAM), gruppo GNFM, for the support. This paper is based upon work from COST Action CA21136  Addressing observational tensions in cosmology with systematics and fundamental physics (CosmoVerse), supported by COST (European Cooperation in Science and Technology). The authors would like to thank Prof. Ravi Kumar Arya, Director of the Xiangshan 5G/6G Laboratory, Zhongshan Institute of Changchun University of Science and Technology, China, for providing access to the HPC facility for the MCMC analysis performed in this work.
\end{acknowledgements}

\bibliographystyle{aa}
\bibliography{mybib}

\clearpage
\appendix
\onecolumn

\section{Additional tables}
\noindent  

Tables~\ref{tab:A1} and~\ref{tab:A2} present the mean values of the cosmological parameters for each model, together with their 68\% (1$\sigma$) confidence levels.

\vspace{0.5em} 

\begin{table*}[!htbp] 
\caption{Numerical values of cosmological parameters at 68\% confidence level for different dataset combinations.}
\label{tab:A1}
\centering
\resizebox{\textwidth}{!}{%
\begin{tabular}{llcccccc}
\hline
Parameter & Dataset & $\Lambda$CDM & o$\Lambda$CDM & $w$CDM & o$w$CDM & CPL \\
\hline
& Ly$\alpha$ + CMB + Galaxy BAO & $0.02234 \pm 0.00012$ & $0.02221 \pm 0.00014$ & $0.02229 \pm 0.00012$ & $0.02219 \pm 0.00014$ & $0.02222 \pm 0.00013$ \\
$\Omega_{b}h^2$ & Ly$\alpha$ + CMB + Pantheon$^{+}$ & $0.02217 \pm 0.00013$ & $0.02221 \pm 0.00013$ & $0.02218 \pm 0.00013$ & $0.02222 \pm 0.00014$ & $0.02220 \pm 0.00013$ \\
& Ly$\alpha$ + CMB + DES-Dovekie & $0.02216 \pm 0.00013$ & $0.02221 \pm 0.00013$ & $0.02218 \pm 0.00013$ & $0.02222 \pm 0.00013$ & $0.02220 \pm 0.00013$ \\
& Ly$\alpha$ + CMB + Union3 & $0.02217 \pm 0.00013$ & $0.02221 \pm 0.00013$ & $0.02218 \pm 0.00013$ & $0.02221 \pm 0.00014$ & $0.02220 \pm 0.00013$ \\
\hline

& Ly$\alpha$ + CMB + Galaxy BAO & $0.11762 \pm 0.00063$ & $0.1194 \pm 0.0011$ & $0.11826 \pm 0.00078$  & $0.1196 \pm 0.0011$ & $0.11940 \pm 0.00084$ \\
$\Omega_{c}h^2$ & Ly$\alpha$ + CMB + Pantheon$^{+}$ & $0.11996 \pm 0.00098$  & $0.1195 \pm 0.0011$ & $0.11977 \pm 0.00097$ & $0.1194 \pm 0.0012$ & $0.1196 \pm 0.0010$ \\
& Ly$\alpha$ + CMB + DES-Dovekie & $0.12005 \pm 0.00095$ & $0.1195 \pm 0.0011$ & $0.1198 \pm 0.0010$ & $0.1194 \pm 0.0012$ & $0.1195 \pm 0.0010$ \\
& Ly$\alpha$ + CMB + Union3 & $0.1200 \pm 0.0010$ & $0.1195 \pm 0.0012$ & $0.1198 \pm 0.0010$ & $0.1194_{-0.0011}^{+0.0012}$ & $0.1195 \pm 0.0010$ \\
\hline

& Ly$\alpha$ + CMB + Galaxy BAO & $1.04102 \pm 0.00024$ & $1.04078 \pm 0.00026$  & $1.04096 \pm 0.00024$ & $1.04079 \pm 0.00025$ & $1.04081 \pm 0.00024$ \\
$100\,\theta_{\mathrm{MC}}$ & Ly$\alpha$ + CMB + Pantheon$^{+}$ & $1.04074 \pm 0.00024$ & $1.04079 \pm 0.00026$ & $1.04076 \pm 0.00025$ & $1.04079 \pm 0.00027$ & $1.04079 \pm 0.00025$ \\
& Ly$\alpha$ + CMB + DES-Dovekie & $1.04073 \pm 0.00025$ & $1.04079 \pm 0.00026$ & $1.04077 \pm 0.00024$ & $1.04080 \pm 0.00025$ & $1.04079 \pm 0.00025$ \\
& Ly$\alpha$ + CMB + Union3 & $1.04074 \pm 0.00025$ & $1.04080 \pm 0.00026$ & $1.04077 \pm 0.00025$ & $1.04079 \pm 0.00026$ & $1.04080 \pm 0.00025$ \\
\hline

& Ly$\alpha$ + CMB + Galaxy BAO & $0.0590 \pm 0.0073$ & $0.0555 \pm 0.0070$ & $0.0565 \pm 0.0069$ & $0.0549 \pm 0.0072$ & $0.0508 \pm 0.0069$ \\
$\tau$ & Ly$\alpha$ + CMB + Pantheon$^{+}$ & $0.0515 \pm 0.0069$  & $0.0513 \pm 0.0072$ & $0.0526 \pm 0.0071$ & $0.0518 \pm 0.0075$ & $0.0519 \pm 0.0071$ \\
& Ly$\alpha$ + CMB + DES-Dovekie & $0.0511 \pm 0.0071$ & $0.0507 \pm 0.0072$ & $0.0529 \pm 0.0073$ & $0.0514_{-0.0074}^{+0.0066}$ & $0.0512 \pm 0.0072$ \\
& Ly$\alpha$ + CMB + Union3 & $0.0513 \pm 0.0071$ & $0.0504 \pm 0.0071$ & $0.0530 \pm 0.0073$ & $0.0519 \pm 0.0073$ & $0.0512 \pm 0.0071$ \\
\hline

& Ly$\alpha$ + CMB + Galaxy BAO & $3.048 \pm 0.014$ & $3.044 \pm 0.014$ & $3.043 \pm 0.014$ & $3.042 \pm 0.014$ &  $3.033 \pm 0.014$ \\
$\ln(10^{10} A_s)$ & Ly$\alpha$ + CMB + Pantheon$^{+}$ & $3.036 \pm 0.013$ & $3.034 \pm 0.014$ & $3.038 \pm 0.014$ & $3.035 \pm 0.015$ & $3.036 \pm 0.014$ \\
& Ly$\alpha$ + CMB + DES-Dovekie & $3.035 \pm 0.014$ & $3.033 \pm 0.015$ & $3.038 \pm 0.014$ & $3.034_{-0.015}^{+0.013}$ & $3.034 \pm 0.014$ \\
& Ly$\alpha$ + CMB + Union3 & $3.036 \pm 0.014$ & $ 3.032 \pm 0.014$ & $3.039 \pm 0.014$ & $3.036 \pm 0.015$ & $3.034 \pm 0.014$ \\
\hline

& Ly$\alpha$ + CMB + Galaxy BAO & $0.9688^{+0.0036}_{-0.0031}$ & $0.9645 \pm 0.0040$ & $0.9673 \pm 0.0035$ & $0.9641 \pm 0.0039$ & $0.9645 \pm 0.0037$ \\
$n_s$ & Ly$\alpha$ + CMB + Pantheon$^{+}$ & $0.9628 \pm 0.0038$ &  $0.9642 \pm 0.0040$ & $0.9636 \pm 0.0038$ & $0.9645 \pm 0.0041$ & $0.9640 \pm0.0041$ \\
& Ly$\alpha$ + CMB + DES-Dovekie & $0.9627 \pm 0.0039$ & $0.9641 \pm 0.0041$ & $0.9636 \pm 0.0039$ & $0.9646 \pm 0.0041$ & $0.9641 \pm 0.0039$ \\
& Ly$\alpha$ + CMB + Union3 & $0.9630 \pm 0.0039$ & $0.9644 \pm 0.0042$ & $0.9634 \pm 0.0040$ & $0.9645 \pm 0.0041$ & $0.9644 \pm 0.0040$ \\
\hline

& Ly$\alpha$ + CMB + Galaxy BAO & $68.19 \pm 0.29$ & $68.48 \pm 0.33$ & $69.30 \pm 0.90$ & $69.09 \pm 0.92$ & $63.8^{+1.7}_{-2.0}$ \\
$H_0$ & Ly$\alpha$ + CMB + Pantheon$^{+}$ & $67.13 \pm 0.43$ & $66.4 \pm 1.1$ & $66.58^{+0.73}_{-0.82}$ & $66.1 \pm 1.1$ & $67.4 \pm 1.1$ \\
& Ly$\alpha$ + CMB + DES-Dovekie & $67.09 \pm 0.43$ & $66.32 \pm 0.98$ & $66.52 \pm 0.71$ & $66.1 \pm 1.0$ & $67.53 \pm 0.88$ \\
& Ly$\alpha$ + CMB + Union3 & $67.12 \pm 0.44$ & $66.2 \pm 1.1$ & $65.5 \pm 1.2$ & $64.8 \pm 1.4$ & $66.3 \pm 1.2$ \\
\hline

& Ly$\alpha$ + CMB + Galaxy BAO & $0.3027 \pm 0.0035$ & $0.3034 \pm 0.0036$ & $0.2927 \pm 0.0072$ & $0.2985 \pm 0.0080$ & $0.353 \pm 0.021$  \\
$\Omega_{m}$ & Ly$\alpha$ + CMB + Pantheon$^{+}$ & $0.3169 \pm 0.0060$ & $0.3230_{-0.011}^{+0.0096}$ & $0.3218 \pm 0.0083$ & $0.326 \pm 0.011$ & $0.314 \pm 0.011$ \\
& Ly$\alpha$ + CMB + DES-Dovekie & $0.3175 \pm 0.0059$ & $0.3239 \pm 0.0094$ & $0.3223^{+0.0072}_{-0.0081}$ & $0.3262 \pm 0.0099$ & $0.3124^{+0.0084}_{-0.0094}$ \\
& Ly$\alpha$ + CMB + Union3 & $0.3171 \pm 0.0061$ & $0.325 \pm 0.011$ & $0.333 \pm 0.013$ & $0.339 \pm 0.015$ & $0.324^{+0.012}_{-0.013}$ \\
\hline

& Ly$\alpha$ + CMB + Galaxy BAO & ---  & $2.2 \pm 1.1$ & --- & $2.0 \pm 1.2$ &  --- \\
$10^{3}\Omega_{k}$ & Ly$\alpha$ + CMB + Pantheon$^{+}$ & --- & $-2.1_{-2.4}^{+2.8}$ & --- & $-1.8 \pm 2.8$ & --- \\
& Ly$\alpha$ + CMB + DES-Dovekie & --- &  $-2.2_{-2.4}^{+2.6}$ & --- & $-1.9 \pm 2.8$ & --- \\
& Ly$\alpha$ + CMB + Union3 & --- & $-2.7_{-2.6}^{+2.9}$  & --- & $-2.4_{-2.7}^{+3.0}$ & --- \\
\hline

& Ly$\alpha$ + CMB + Galaxy BAO & --- & --- & $-1.046 \pm 0.036$ & $-1.026 \pm 0.038$ & $-0.42 \pm 0.21$  \\
$w$ or $w_0$ & Ly$\alpha$ + CMB + Pantheon$^{+}$ & --- & --- & $-0.979 \pm 0.027$ & $-0.982 \pm 0.029$ & $-0.874 \pm 0.076$ \\
& Ly$\alpha$ + CMB + DES-Dovekie & --- & --- & $-0.977 \pm 0.024$ & $-0.984 \pm 0.026$ & $-0.826 \pm 0.093$ \\
& Ly$\alpha$ + CMB + Union3 & --- & --- & $-0.942 \pm 0.039$ & $-0.946 \pm 0.039$ & $-0.70^{+0.12}_{-0.14}$ \\
\hline

& Ly$\alpha$ + CMB + Galaxy BAO & --- & --- & --- & --- & $-1.72 \pm 0.58$ \\
$w_a$ & Ly$\alpha$ + CMB + Pantheon$^{+}$ & ---  & --- & --- & --- & $-0.53^{+0.42}_{-0.33}$ \\
& Ly$\alpha$ + CMB + DES-Dovekie & --- & --- & --- & --- & $-0.71^{+0.47}_{-0.39}$ \\
& Ly$\alpha$ + CMB + Union3 & --- & --- & --- & --- & $-1.06^{+0.58}_{-0.44}$ \\
\hline

& Ly$\alpha$ + CMB + Galaxy BAO & --- & -4.82 & -1.57 & -4.40 & -12.48 \\
$\Delta{\chi^{2}_{\text{MAP}}}$ & Ly$\alpha$ + CMB + Pantheon$^{+}$ & --- & -0.40 & -0.60 & -1.25 & -1.08 \\
& Ly$\alpha$ + CMB + DES-Dovekie & --- & 0.24 &  -0.16 & 0.63 & -1.64 \\
& Ly$\alpha$ + CMB + Union3 & --- & -1.20 & -2.75 & -2.79 & -6.28  \\ \hline

& Ly$\alpha$ + CMB + Galaxy BAO & --- & 2.20 & 1.25 & 1.59 & 3.10 \\
 $N\sigma$ & Ly$\alpha$ + CMB + Pantheon$^{+}$ & --- & 0.63 & 0.77 & 0.62 & 0.55 \\
& Ly$\alpha$ + CMB + DES-Dovekie & --- & 0.00 & 0.40 & 0.00 & 0.77 \\
& Ly$\alpha$ + CMB + Union3 & --- & 1.10 & 1.66 & 1.16 & 2.02  \\ \hline

& Ly$\alpha$ + CMB + Galaxy BAO & 0 & 1.47 & 3.32 & 4.17 & 1.39 \\
 $|\ln B_{i,j}|$ & Ly$\alpha$ + CMB + Pantheon$^{+}$ & 0 & 3.80 & 3.36 & 5.94 & 4.22 \\
& Ly$\alpha$ + CMB + DES-Dovekie & 0 & 3.80 & 3.61 & 5.93 & 3.07 \\
& Ly$\alpha$ + CMB + Union3 & 0 & 3.70 & 2.46 & 4.91 & 1.53 \\ 
\hline
\end{tabular}
}
\begin{flushleft}
\textit{Notes.} Constraints are given at 68\% (1$\sigma$) confidence level for the $\Lambda$CDM, o$\Lambda$CDM, $w$CDM, o$w$CDM, and CPL using different combinations of DESI DR2 BAO datasets with the CMB and SNe Ia (Pantheon$^+$, DES-Dovekie, and Union3) samples.
\end{flushleft}
\end{table*}

\begin{table*}
\caption{Numerical values of cosmological parameters at 68\% confidence level for different dataset combinations.}
\label{tab:A2}
\centering
\resizebox{\textwidth}{!}{%
\begin{tabular}{llcccccc}
\hline
Parameter & Dataset & Logarithmic & Exponential & JBP & BA & GEDE \\ 
\hline
& Ly$\alpha$ + CMB + Galaxy BAO & $0.02221 \pm 0.00013$ & $0.02221 \pm 0.00012$ & $0.02227 \pm 0.00013$ & $0.02221 \pm 0.00012$ & $0.02228 \pm 0.00013$ \\
$\Omega_{b}h^2$ & Ly$\alpha$ + CMB + Pantheon$^{+}$ & $0.02220 \pm 0.00013$ & $0.02220 \pm 0.00013$ & $0.02219 \pm 0.00013$ & $0.02220 \pm 0.00013$ & $0.02219 \pm 0.00013$ \\
& Ly$\alpha$ + CMB + DES-Dovekie & $0.02221 \pm 0.00013$ & $0.02221 \pm 0.00013$ & $0.02219 \pm 0.00013$ & $0.02220 \pm 0.00013$ & $0.02219 \pm 0.00013$ \\
& Ly$\alpha$ + CMB + Union3 & $0.02220 \pm 0.00013$ & $0.02221 \pm 0.00013$ & $0.02219 \pm 0.00013$ & $0.02220 \pm 0.00013$ & $0.02218 \pm 0.00013$ \\
\hline

& Ly$\alpha$ + CMB + Galaxy BAO & $0.11941 \pm 0.00085$ & $0.11942 \pm 0.00084$ & $0.11850 \pm 0.00078$ & $0.11940 \pm 0.00083$ & $0.11836 \pm 0.00083$ \\
$\Omega_{c}h^2$ & Ly$\alpha$ + CMB + Pantheon$^{+}$ & $0.1195 \pm 0.0010$ & $0.1196 \pm 0.0010$ & $0.1197 \pm 0.0010$  & $0.1196 \pm 0.0010$ & $0.11978 \pm 0.00099$ \\
& Ly$\alpha$ + CMB + DES-Dovekie & $0.1195 \pm 0.0010$ & $0.11948 \pm 0.00099$ & $0.1197 \pm 0.0010$ & $0.1195 \pm 0.0010$ & $0.1197 \pm 0.0010$ \\
& Ly$\alpha$ + CMB + Union3 & $0.1195 \pm 0.0010$ & $0.1195 \pm 0.0010$ & $0.1197 \pm 0.0011$ & $0.1195 \pm 0.0010$ & $0.1198 \pm 0.0010$ \\
\hline

& Ly$\alpha$ + CMB + Galaxy BAO & $1.04080 \pm 0.00024$ & $1.04080 \pm 0.00023$ & $1.04092 \pm 0.00023$ & $1.04081 \pm 0.00024$ & $1.04093 \pm 0.00024$ \\
$100\,\theta_{\mathrm{MC}}$ & Ly$\alpha$ + CMB + Pantheon$^{+}$ & $1.04079 \pm 0.00025$ & $1.04079 \pm 0.00025$ & $1.04077 \pm 0.00026$ & $1.04079 \pm 0.00025$ & $1.04076 \pm 0.00026$ \\
& Ly$\alpha$ + CMB + DES-Dovekie & $1.04079 \pm 0.00025$ & $1.04080 \pm 0.00025$ & $1.04077 \pm 0.00024$ & $1.04078 \pm 0.00025$ & $1.04077 \pm 0.00025$ \\
& Ly$\alpha$ + CMB + Union3 & $1.04080 \pm 0.00026$ & $1.04080 \pm 0.00025$ & $1.04077 \pm 0.00025$ & $1.04079 \pm 0.00025$ & $1.04076 \pm 0.00025$ \\
\hline

& Ly$\alpha$ + CMB + Galaxy BAO & $0.0507 \pm 0.0071$ & $0.0509 \pm 0.0070$ & $0.0557 \pm 0.0069$ & $0.0508 \pm 0.0073$ & $0.0560 \pm 0.0071$ \\
$\tau$ & Ly$\alpha$ + CMB + Pantheon$^{+}$ & $0.0515 \pm 0.0074$ & $0.0514 \pm 0.0073$ & $0.0522 \pm 0.0073$ & $0.0518 \pm 0.0073$ & $0.0526 \pm 0.0071$ \\
& Ly$\alpha$ + CMB + DES-Dovekie & $0.0513 \pm 0.0072$ & $0.0511 \pm 0.0071$ & $0.0516 \pm 0.0071$ & $0.0514 \pm 0.0073$ & $0.0527 \pm 0.0072$ \\
& Ly$\alpha$ + CMB + Union3 & $0.0509 \pm 0.0073$ & $0.0511 \pm 0.0071$ & $0.0518 \pm 0.0073$ & $0.0509 \pm 0.0070$ & $0.0528 \pm 0.0069$ \\
\hline

& Ly$\alpha$ + CMB + Galaxy BAO & $3.033 \pm 0.014$ & $3.033 \pm 0.014$ & $3.042 \pm 0.014$ & $3.033 \pm 0.014$ & $3.042 \pm 0.014$ \\
$\ln(10^{10} A_s)$ & Ly$\alpha$ + CMB + Pantheon$^{+}$ & $3.035 \pm 0.014$ & $3.035 \pm 0.014$ & $3.037 \pm 0.014$ & $3.036 \pm 0.014$ & $3.038 \pm 0.014$ \\
& Ly$\alpha$ + CMB + DES-Dovekie & $3.034 \pm 0.014$ & $ 3.034 \pm 0.014$ & $3.035 \pm 0.014$ & $3.035 \pm 0.014$ & $3.038 \pm 0.014$ \\
& Ly$\alpha$ + CMB + Union3 & $3.033 \pm 0.014$ & $3.034 \pm 0.014$ & $3.036 \pm 0.014$ & $3.033 \pm 0.014$ & $3.039 \pm 0.014$ \\
\hline

& Ly$\alpha$ + CMB + Galaxy BAO & $0.9646 \pm 0.0038$ & $0.9644 \pm 0.0037$ & $0.9665 \pm 0.0036$ & $0.9644 \pm 0.0035$ & $0.9669 \pm 0.0036$ \\
$n_s$ & Ly$\alpha$ + CMB + Pantheon$^{+}$ & $0.9641_{-0.0038}^{+0.0044}$ & $0.9640_{-0.0037}^{+0.0041}$  & $0.9636 \pm 0.0040$ & $0.9641 \pm 0.0038$ & $0.9634 \pm 0.0039$ \\
& Ly$\alpha$ + CMB + DES-Dovekie & $0.9644 \pm 0.0040$ & $0.9640 \pm 0.0039$ & $0.9637 \pm 0.0040$ & $0.9643 \pm 0.0039$ & $0.9638 \pm 0.0039$ \\
& Ly$\alpha$ + CMB + Union3 & $0.9643 \pm 0.0039$ & $0.9644 \pm 0.0039$ & $0.9636 \pm 0.0040$ & $0.9642 \pm 0.0039$ & $0.9633 \pm 0.0039$ \\
\hline

& Ly$\alpha$ + CMB + Galaxy BAO & $64.6 \pm 1.8$ & $64.3 \pm 1.9$ & $66.35_{-1.9}^{+0.99}$ & $63.8 \pm 1.8$ & $69.39 \pm 0.91$ \\
$H_0$ & Ly$\alpha$ + CMB + Pantheon$^{+}$ & $67.5 \pm 1.0$ & $67.4 \pm 1.1$ & $67.00 \pm 0.95$ & $67.3 \pm 1.1$ & $66.60 \pm 0.79$  \\
& Ly$\alpha$ + CMB + DES-Dovekie & $67.62 \pm 0.91$ & $67.62 \pm 0.93$ & $67.17 \pm 0.78$ & $67.56 \pm 0.91$ & $66.55 \pm 0.70$ \\
& Ly$\alpha$ + CMB + Union3 & $66.4 \pm 1.3$ & $66.5 \pm 1.3$ & $65.8 \pm 1.2$ & $66.5 \pm 1.2$ & $65.5 \pm 1.2$ \\
\hline

& Ly$\alpha$ + CMB + Galaxy BAO & $0.341^{+0.018}_{-0.021}$ & $0.345^{+0.019}_{-0.023}$ & $0.322_{-0.010}^{+0.018}$ & $0.351^{+0.019}_{-0.022}$ & $0.2935 \pm 0.0072$ \\
$\Omega_{m0}$ & Ly$\alpha$ + CMB + Pantheon$^{+}$ & $0.313 \pm 0.010$ & $0.314 \pm 0.011$ & $0.3177 \pm 0.0098$ & $0.314 \pm 0.011$ & $0.3217 \pm 0.0084$  \\
& Ly$\alpha$ + CMB + DES-Dovekie & $0.3115 \pm 0.0092$ & $0.3114 \pm 0.0093$ & $0.3160 \pm 0.0081$ & $0.3121^{+0.0086}_{-0.0097}$ & $0.3219 \pm 0.0075$ \\
& Ly$\alpha$ + CMB + Union3 & $0.323 \pm 0.013$ & $0.323^{+0.012}_{-0.014}$ & $0.329 \pm 0.013$ & $0.322^{+0.012}_{-0.013}$ & $0.333 \pm 0.013$ \\
\hline

& Ly$\alpha$ + CMB + Galaxy BAO & $-0.58_{-0.20}^{+0.18}$ & $-0.54_{-0.21}^{+0.17}$  & $-0.640_{-0.078}^{+0.23}$ & $-0.51_{-0.19}^{+0.17}$ & --- \\
$w$ or $w_0$ & Ly$\alpha$ + CMB + Pantheon$^{+}$ & $-0.879_{-0.074}^{+0.061}$ & $-0.898 \pm 0.076$ & $-0.90 \pm 0.11$ & $-0.907 \pm 0.073$ & --- \\
& Ly$\alpha$ + CMB + DES-Dovekie & $-0.847_{-0.089}^{+0.074}$ & $-0.839 \pm 0.086$ & $-0.77_{-0.10}^{+0.13}$ & $-0.861 \pm 0.081$ & --- \\
& Ly$\alpha$ + CMB + Union3 & $-0.75_{-0.12}^{+0.10}$ & $-0.75_{-0.12}^{+0.10}$ & $-0.68_{-0.10}^{+0.16}$ & $-0.76 \pm 0.11$ & --- \\
\hline

& Ly$\alpha$ + CMB + Galaxy BAO & $-0.98_{-0.32}^{+0.45}$ & $-1.13_{-0.58}^{+0.51}$  & $<-1.71$ & $-0.81_{-0.25}^{+0.31}$ & --- \\
$w_a$ & Ly$\alpha$ + CMB + Pantheon$^{+}$ & $-0.33_{-0.19}^{+0.31}$ & $-0.34_{-0.26}^{+0.33}$  &  $-0.60 \pm 0.78$ & $-0.21_{-0.19}^{+0.22}$ & --- \\
& Ly$\alpha$ + CMB + DES-Dovekie & $-0.50_{-0.26}^{+0.35}$ & $-0.55_{-0.31}^{+0.37}$ & $-1.45_{-0.90}^{+0.69}$ & $-0.33_{-0.20}^{+0.24}$ & --- \\
& Ly$\alpha$ + CMB + Union3 & $-0.66_{-0.30}^{+0.46}$ & $-0.72_{-0.34}^{+0.47}$ & $<-1.34$ & $-0.48_{-0.23}^{+0.31}$ & --- \\
\hline

& Ly$\alpha$ + CMB + Galaxy BAO & --- & --- & --- & --- & $0.36 \pm 0.26$ \\
$\Delta$ & Ly$\alpha$ + CMB + Pantheon$^{+}$ & --- & --- & --- & --- & $-0.14 \pm 0.19$ \\
& Ly$\alpha$ + CMB + DES-Dovekie & --- & --- & --- & --- & $-0.16 \pm 0.17$ \\
& Ly$\alpha$ + CMB + Union3 & --- & --- & --- & --- & $-0.40 \pm 0.27$ \\
\hline

& Ly$\alpha$ + CMB + Galaxy BAO & -9.63 & -9.98 &  -6.67 & -10.74 &  -2.95 \\
$\Delta{\chi^{2}_{\text{MAP}}}$ & Ly$\alpha$ + CMB + Pantheon$^{+}$ & -1.45 & -1.29 & -1.76 & -0.83 & -1.18 \\
& Ly$\alpha$ + CMB + DES-Dovekie & -0.94  & -1.71 & -2.97 & -0.99 & 0.15 \\
& Ly$\alpha$ + CMB + Union3 & -4.02 & -4.98 & -4.58 & -5.22 & -1.34  \\
\hline

& Ly$\alpha$ + CMB + Galaxy BAO & 2.65 & 2.71 & 2.10 & 2.83 & 1.72 \\
 $N\sigma$ & Ly$\alpha$ + CMB + Pantheon$^{+}$ & 0.70 & 0.64 & 0.82 & 0.44 & 1.09 \\
& Ly$\alpha$ + CMB + DES-Dovekie & 0.49 & 0.80 & 1.21 & 0.51 & 0.00 \\
& Ly$\alpha$ + CMB + Union3 & 1.50 & 1.73 & 1.64 & 1.79 & 1.16  \\
\hline

& Ly$\alpha$ + CMB + Galaxy BAO & 0.59 & 0.17 & 1.87 & 0.58 & 1.77 \\
 $|\ln B_{i,j}|$ & Ly$\alpha$ + CMB + Pantheon$^{+}$ & 4.58 & 4.90 & 4.19 & 4.62 & 2.93 \\
& Ly$\alpha$ + CMB + DES-Dovekie & 3.40 & 3.91 & 2.38 & 3.76 & 2.66 \\
& Ly$\alpha$ + CMB + Union3 & 1.99 & 2.09 & 1.50 & 2.24 & 2.05 \\ 
\hline
\end{tabular}
}
\begin{flushleft}
\textit{Notes.} Constraints are given at 68\% (1$\sigma$) confidence level for the Logarithmic, Exponential, JBP, BA, and GEDE using different combinations of DESI DR2 BAO datasets with the CMB and SNe Ia (Pantheon$^+$, DES-Dovekie, and Union3) samples.
\end{flushleft}
\end{table*}

\end{document}